\documentclass[fleqn,10pt]{wlscirep}
\usepackage[utf8]{inputenc}
\usepackage[T1]{fontenc}
\usepackage{lineno}
\usepackage{amsmath} 
\usepackage{multirow}
\usepackage{amsmath}
\usepackage{graphicx} 

\title{Towards a Comprehensive Benchmark for Pathological Lymph Node Metastasis in Breast Cancer Sections}

\author[1,$\dag$]{Xitong Ling}
\author[2,$\dag$]{Yuanyuan Lei}
\author[1,$\dag$]{Jiawen Li}
\author[3,]{Junru Cheng}
\author[2,]{Wenting Huang}
\author[1]{Tian Guan}
\author[2,*]{Jian Guan}
\author[1,*]{Yonghong He}
\affil[1]{Shenzhen International Graduate School, Tsinghua University, Shenzhen, 518071, China}
\affil[2]{National Cancer Center/National Clinical Research Center for Cancer/Cancer Hospital \& Shenzhen Hospital, Chinese Academy of Medical Sciences and Peking Union Medical College, Shenzhen, 518116, China}
\affil[3]{Research Institute of Tsinghua, Guangzhou, 508157, China}

\affil[*]{corresponding author(s): Yonghong He (heyh@sz.tsinghua.edu.cn), Jian Guan (graceguan9@hotmail.com) }

\affil[$\dag$]{these authors contributed equally to this work}

\begin{abstract}

Advances in optical microscopy scanning have significantly contributed to computational pathology (CPath) by converting traditional histopathological slides into whole slide images (WSIs). This development enables comprehensive digital reviews by pathologists and accelerates AI-driven diagnostic support for WSI analysis. Recent advances in foundational pathology models have increased the need for benchmarking tasks. The Camelyon series~\cite{bejnordi2017diagnostic,bandi2018detection} is one of the most widely used open-source datasets in computational pathology. However, the quality, accessibility, and clinical relevance of the labels have not been comprehensively evaluated.In this study, we reprocessed 1,399 WSIs and labels from the Camelyon-16~\cite{bejnordi2017diagnostic} and Camelyon-17~\cite{bandi2018detection} datasets, removing low-quality slides, correcting erroneous labels, and providing expert pixel annotations for tumor regions in the previously unreleased test set. Based on the sizes of re-annotated tumor regions, we upgraded the binary cancer screening task to a four-class task: negative, micro-metastasis, macro-metastasis, and Isolated Tumor Cells (ITC). 
We reevaluated pre-trained pathology feature extractors and multiple instance learning (MIL) methods using the cleaned dataset, providing a benchmark that advances AI development in histopathology.

\end{abstract}
\begin{document}

\flushbottom
\maketitle
\thispagestyle{empty}   
\section*{Background \& Summary}

The efficient utilization of digital pathology and computational resources has led to the rapid rise of AI-based computational pathology~\cite{song2023artificial,van2021deep}. In recent years, general foundational models for pathology, pre-trained on large-scale data, have garnered significant attention~\cite{huang2023visual,chen2024towards,xu2024whole,wang2022transformer,lu2024visual}. These models have demonstrated strong feature extraction capabilities for pathological images, as evidenced by evaluations across a series of whole-slide image-level downstream tasks~\cite{kather2019predicting,pataki2022huncrc,barbano2021unitopatho}.
For example, CTransPath~\cite{wang2022transformer} uses a Semantically-Relevant Contrastive Learning (SRCL) framework to pre-train a CNN-Transformer hybrid feature extractor on 150 million patches, with its effectiveness validated across five downstream tasks.
UNI\cite{chen2024towards} employed the self-supervised DINO-v2\cite{oquab2023dinov2} method to train a robust general pathology visual encoder on one billion patches from approximately 100,000 whole slide images (WSIs). Gigapath\cite{xu2024whole} utilized 1.3 billion patches to train a LongNet\cite{ding2023longnet} architecture-based visual encoder for slide-level representation learning. These pathology-pre-trained models have demonstrated superior performance in downstream tasks including tumor classification, survival analysis, and lesion segmentation.
PLIP\cite{huang2023visual}, pre-trained on approximately 200,000 pathology image-text pairs collected from medical Twitter, developed a multimodal pathology foundational model using contrastive learning~\cite{radford2021learning}, capable of both image and text comprehension. 
CONCH~\cite{lu2024visual} employs CoCa~\cite{yu2022coca} for self-supervised pre-training on 1.17 million image–caption pairs and has been extensively evaluated across 14 downstream benchmarks, demonstrating its outstanding performance.

Acquiring finely annotated large-scale pathology image datasets remains challenging due to the extremely high resolution of pathology images and the specialized expertise required for annotations. Nonetheless, the continued development of foundational models and downstream tasks in computational pathology makes high-quality pathology image datasets increasingly essential.

\begin{figure}[htbp]
\centering
\includegraphics[width=0.96\linewidth]{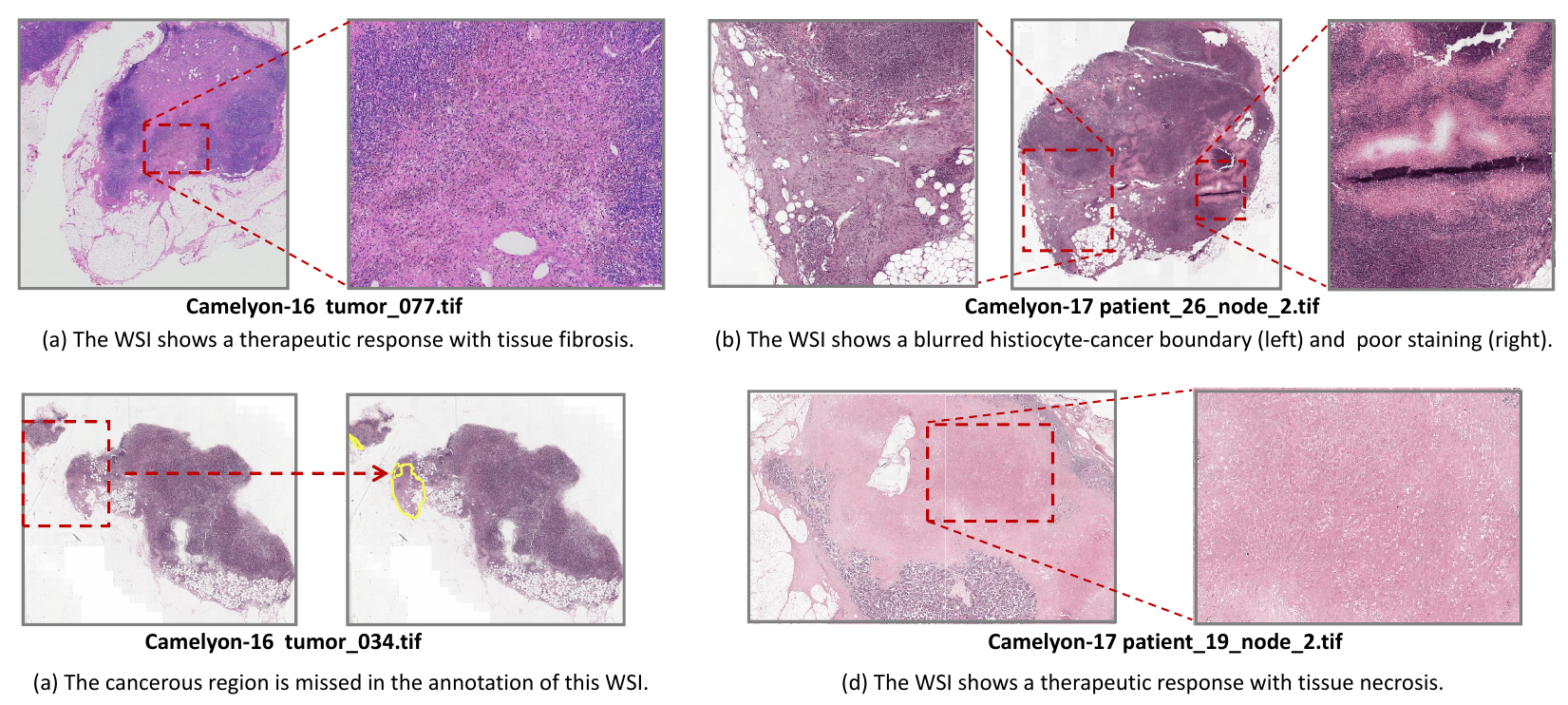}
\caption{Issues within the Camelyon-16~\cite{bejnordi2017diagnostic} and Camelyon-17~\cite{bandi2018detection} datasets: therapeutic response, annotation omissions, blurred boundaries, and poor staining.}
\label{DS-problem}
\end{figure}

The Camelyon series~\cite{bejnordi2017diagnostic,bandi2018detection}(\hyperlink{http://gigadb.org/dataset/100439}{http://gigadb.org/dataset/100439}) , a publicly available pathology dataset focused on detecting breast cancer lymph node metastasis, is widely used for evaluating multiple instance Learning (MIL) methods. However, as shown in Figure~\ref{DS-problem}, some images in the Camelyon series are of poor quality, exhibit treatment-related artifacts, and contain errors in slide-level labeling. The Camelyon-16~\cite{bejnordi2017diagnostic} dataset includes only tumor and negative labels, making it incompatible with Camelyon-17~\cite{bandi2018detection} labels. Many pixel-level annotations are inaccurate, and some slides lack pixel-level annotations entirely. 
These issues hinder the accurate evaluation of deep learning methods in downstream pathology tasks.

In this paper, we filtered out and removed slides from the Camelyon dataset that were blurred, poorly stained, exhibited treatment-related artifacts, or were ambiguous in terms of positivity. Additionally, we expanded the binary classification labels in Camelyon-16~\cite{bejnordi2017diagnostic} to a four-class system to facilitate the merging of Camelyon-16 and Camelyon-17~\cite{bandi2018detection} datasets. Finally, we corrected the pixel-level annotations in the Camelyon dataset and added pixel-level annotations to positive slides that previously lacked them. Using the corrected dataset, we reevaluated 12 mainstream MIL methods, including ABMIL\cite{ilse2018attention}, TransMIL\cite{shao2021transmil} and ClAM\cite{lu2021data}, etc. on two natural image pre-trained feature encoders, ResNet-50~\cite{he2016deep} and VIT-S~\cite{ilse2018attention}, as well as four pathology-specific pre-trained feature encoders, PILP\cite{huang2023visual}, CONCH~\cite{lu2024visual}, UNI\cite{chen2024towards} and Gigapath\cite{xu2024whole}.

\section*{Methods}

\subsection*{Dataset Overview}
The official Camelyon-16~\cite{bejnordi2017diagnostic} dataset contains 399 WSIs, split into 270 for training and 129 for testing. The training set includes 111 tumor slides and 259 negative slides, while the test set includes 49 tumor slides and 80 negative slides. The official Camelyon-17~\cite{bandi2018detection} dataset consists of 1000 WSIs, evenly divided into 500 for training and 500 for testing. The training set consists of 318 negative slides, 59 micro-metastasis slides, 87 macro-metastasis slides, and 36 Isolated Tumor Cells (ITC) slides. The test set labels are not publicly available.
After data cleaning by professional pathologists, the Camelyon-16 dataset consists of 386 WSIs: 238 negative, 71 micro-metastasis, 69 macro-metastasis, and 8 ITC WSIs. The Camelyon-17 dataset consists of 964 WSIs: 633 negative, 103 micro-metastasis, 182 macro-metastasis, and 46 ITC WSIs.
We combined the updated Camelyon-16 and Camelyon-17 datasets to form the Camelyon\(^+\) dataset. It consists of 1,350 WSIs: 871 negative, 174 micro-metastasis, 251 macro-metastasis, and 54 ITC WSIs.

\subsection*{Exclusion Criteria}
We excluded certain WSIs based on the following criteria: focal blurriness, poor staining quality, difficulty distinguishing positive foci, and the presence of treatment-related artifacts.
Of the 49 slides we remove, 26 show therapeutic response, 3 have staining issues, 12 exhibit focal blurring, 4 are of poor quality, and 4 contain suspicious cancerous regions
The presence of treatment response may interfere with model construction. In pathology, tumor treatment response refers to the histological changes in tumors following treatments such as surgery, chemotherapy, radiotherapy, targeted therapy, or immunotherapy, and the corresponding reaction to these treatments. Pathological analysis can assess histological indicators such as tumor cell necrosis, proliferation, and apoptosis, thereby evaluating treatment efficacy. Two typical treatment responses are tissue necrosis and fibrosis. Necrosis refers to areas of dead tissue formed after tumor cells die following treatment. Fibrosis refers to the scar tissue formed as a result of the tumor's self-repair after damage. Necrotic and fibrotic areas can affect the feature representation of tumor regions in computational pathology, thereby impacting the performance of downstream tasks.

\section*{Data Records}

The Camelyon\(^+\) dataset is available via ScienceDB: (\hyperlink{https://doi.org/10.57760/sciencedb.16442}{https://doi.org/10.57760/sciencedb.16442}).
The original WSI data can be downloaded from the official websites of Camelyon16~\cite{bejnordi2017diagnostic} and Camelyon-17~\cite{bandi2018detection}, so it has not been uploaded to the database.
Slide-level labels are recorded in XLSX files.
We provide the classification labels of the merged Camelyon\(^+\) dataset, which combines corrected versions of Camelyon-16 and Camelyon-17. The dataset includes four classification labels (negative, micro, macro, ITC) and two classification labels (negative, tumor), facilitating various downstream tasks.
Since the original training dataset from Camelyon-16 is named using "tumor," "normal," and ID, we have renamed it to prevent pathologists correcting the data from forming prior judgments. The correspondence with the original naming will also be recorded and made public using a XLSX file.
For positive WSIs, pixel-level annotations are provided in XML format.
In order to facilitate future comparative experiments of multiple feature extractors on the Camelyon\(^+\) dataset, we also release the feature files extracted at 20X magnification using ResNet-50\cite{he2016deep}, VIT-S\cite{dosovitskiy2020image}, PLIP\cite{huang2023visual}, CONCH\cite{lu2024visual}, UNI\cite{chen2024towards}, and Gigapath\cite{xu2024whole}.
The feature files are stored in PT format, which can be easily processed using the PyTorch library.
Detailed dataset information and download links are available at:(\hyperlink{https://github.com/lingxitong/CAMELYON-PLUS-BENCHMARK}{https://github.com/lingxitong/CAMELYON-PLUS-BENCHMARK}).

\section*{Technical Validation}

\subsection*{Methodology}
The objective of our designed benchmark is to utilize slide-level labels to predict metastasis types. The commonly used approach is to adopt a deep learning strategy based on MIL, which has been recognized in recent studies for its strong capability to represent slide-level features.~\cite{yan2024shapley,ouyang2024mergeup,chu2024retmil,qiehe2024nciemil,yang2024mambamil}.
MIL is a weakly supervised approach where a single WSI is treated as a bag, and each patch within the WSI is considered an instance. If any instance is cancerous, the entire WSI is labeled as cancerous, while a WSI is classified as normal only if all instances are normal.

With the advancement of deep neural networks, embedding-based MIL has become the dominant approach for WSI analysis. In embedding-based MIL, a pre-trained feature extractor first extracts features from the WSI, followed by an aggregator that pools the features for downstream classification tasks. Mean-MIL and Max-MIL aggregate features using mean pooling and max pooling, respectively, though the pooling mechanism inevitably results in information loss. ABMIL\cite{ilse2018attention} introduces the attention mechanism into MIL, dynamically assigning weights to each instance based on attention scores. CLAM\cite{lu2021data} further enhances this by incorporating instance-level clustering mechanisms to introduce domain knowledge, providing additional supervision alongside attention-based weight assignments. TransMIL\cite{shao2021transmil} leverages self-attention within the MIL aggregator to capture relationships between different instances, thereby improving global modeling capabilities. AMD-MIL\cite{ling2024agent} introduces an agent mechanism into the MIL aggregator and employs threshold filtering for feature selection, improving MIL performance. DSMIL\cite{li2021dual} models instance relationships directly using a dual-stream architecture and a trainable distance measurement module. DTFD\cite{zhang2022dtfd} addresses the issue of limited WSIs by creating pseudo-bags. WiKG\cite{li2024dynamic} treats WSIs as knowledge graphs, dynamically constructing neighboring nodes and directed edges based on relationships between instances, and then updates the head node using knowledge-aware attention. FR-MIL\cite{10640165} introduces a distribution re-calibration approach that adjusts the feature distribution of a WSI bag (instances) based on the statistics of the max-instance (key) feature.

\subsection*{Data Preprocessing}
For all datasets, we crop non-overlapping $256 \times 256$ patches at $20 \scalebox{1.3}{$\times$}$ magnification. 
We then use six feature extractors ResNet-50\cite{he2016deep}, VIT-S\cite{dosovitskiy2020image}, PLIP\cite{huang2023visual}, CONCH\cite{lu2024visual}, UNI\cite{chen2024towards}, and Gigapath\cite{xu2024whole} to extract features from the WSIs. Subsequently, we conduct two sets of experiments. The first set is a comparative experiment on the Camelyon-17~\cite{bandi2018detection} dataset before and after label correction. The Camelyon-17-Origin dataset follows the official split, with 500 WSIs for training and 500 WSIs for testing. The Camelyon-17-Refine dataset also maintains the official split but excludes slides that fall under exclusion criteria. The Camelyon-17-Refine training set contains 492 slides, while the test set includes 472 slides. This comparative experiment evaluates the impact of dataset quality on MIL models. Since the original version of Camelyon-16~\cite{bejnordi2017diagnostic} does not have four-class labels, we do not perform similar experiments on it. The next set is the benchmark experiments on Camelyon\(^+\), we evaluate using five-fold cross-validation, with each fold employing stratified sampling to maintain a fixed proportion of different classes. Since each patient has multiple slides in the Camelyon-17 dataset, in order to prevent data leakage, we ensure that slides of the same label of the same patient do not appear in the training set and the validation set at the same time.

\subsection*{Camelyon-17 Comparative Experiment}
\begin{figure}[htbp]
\centering
\includegraphics[width=0.96\linewidth]{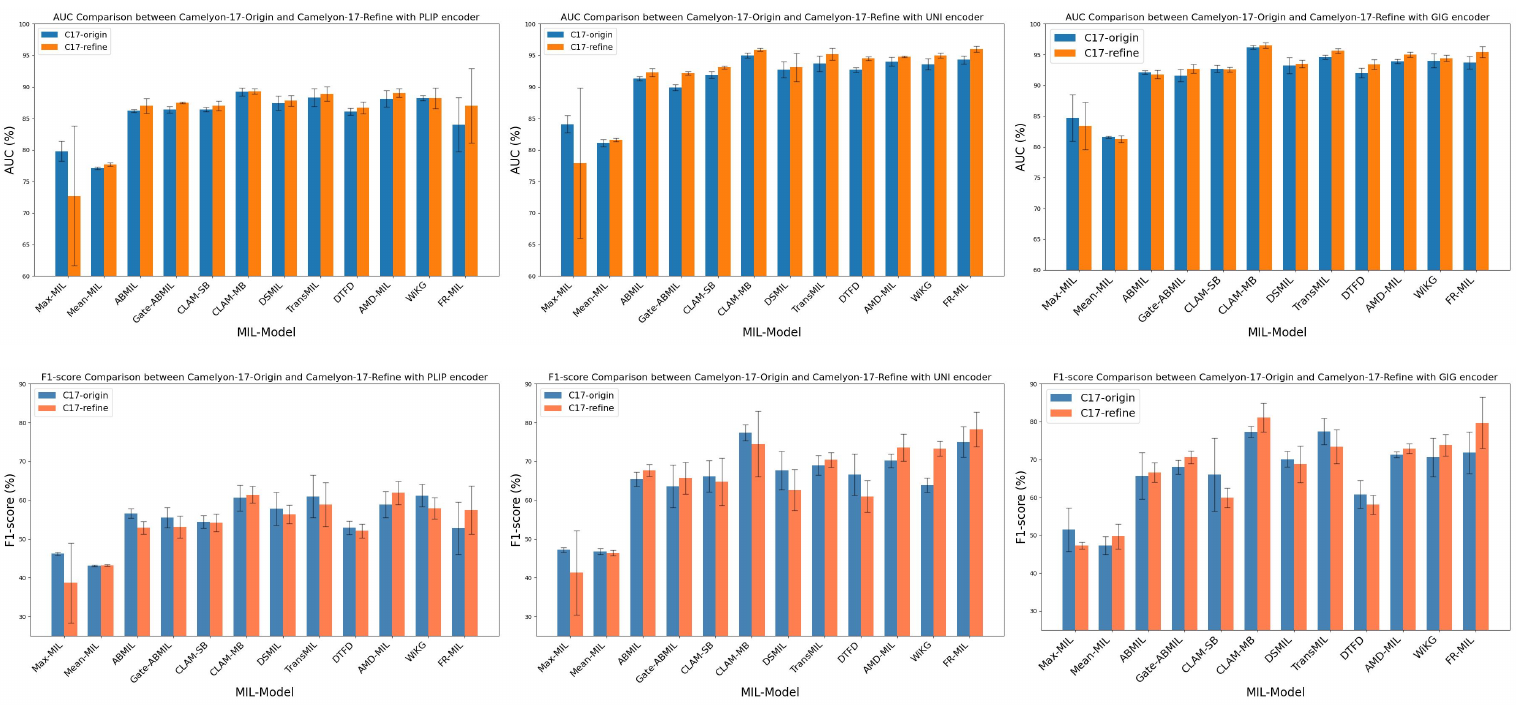}
\caption{AUC and F1-score Comparison of Different Methods in the Camelyon-17-Origin and Camelyon-17-Refine Comparative Experiments.}
\label{C17-Com}
\end{figure}

\begin{figure}[htbp]
\centering
\includegraphics[width=0.96\linewidth]{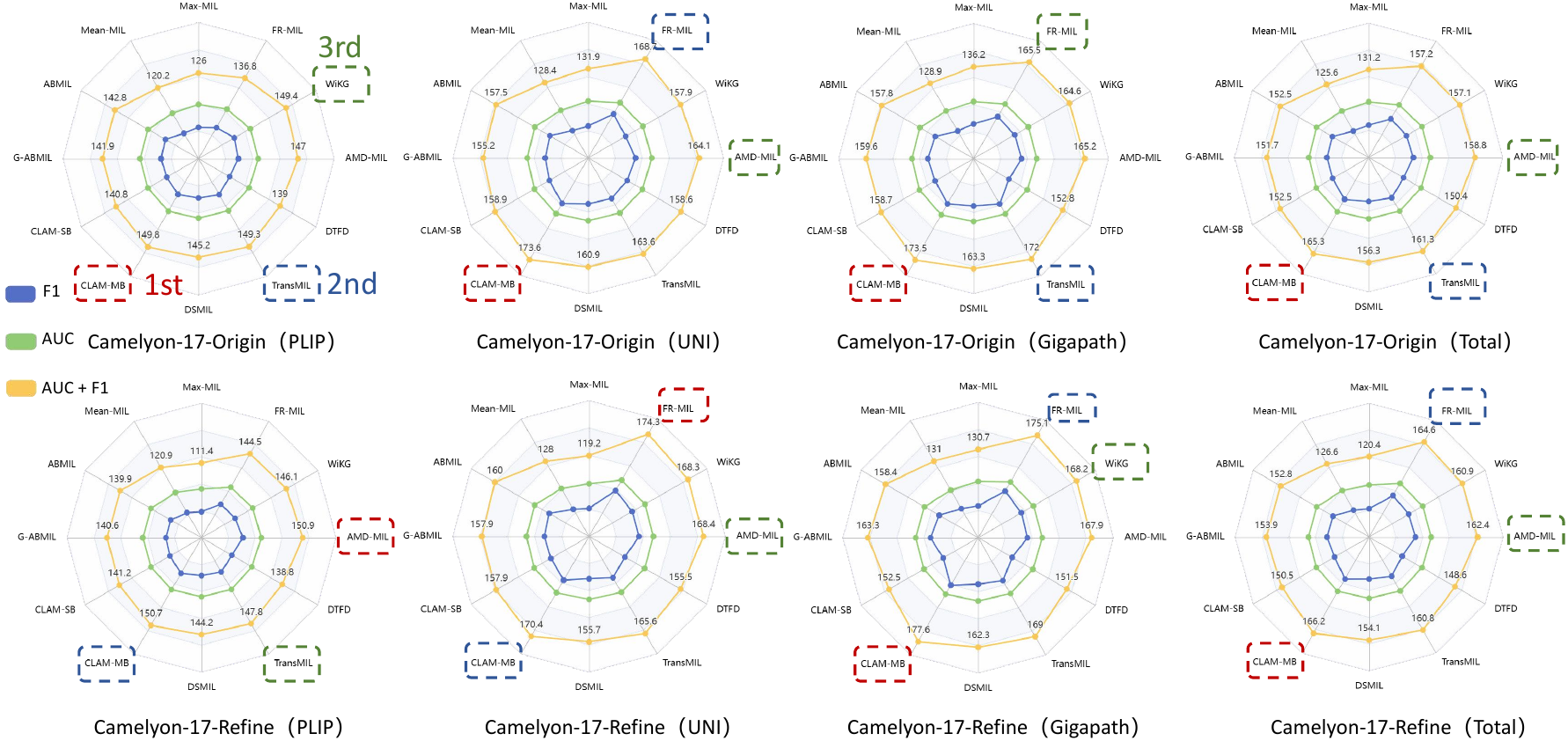}
\caption{Model ranking radar chart based on AUC and F1-score on Camelyon-17-Origin and Camelyon-17-Refine datasets.}
\label{Model-rank}
\end{figure}

In the comparative experiments before and after correction on the Camelyon-17~\cite{bandi2018detection} dataset, we primarily evaluated three pathology pre-trained feature extractors: PLIP\cite{huang2023visual}, UNI\cite{chen2024towards}, and Gigapath\cite{xu2024whole}. The learning rate was set to 2e-4, using the Adam optimizer with a weight decay of 1e-5. We repeated the experiments with random seeds of 2023, 2024, and 2025, and reported the mean and standard deviation of the evaluation metrics as shown in Table~\ref{C17-Origin-Com} and Table~\ref{C17-refine-Com}. All experiments were conducted on a workstation equipped with 4 NVIDIA RTX 3090 GPUs.
Due to the significant class imbalance in the Camelyon-17 four-class dataset, our analysis concentrated on two key evaluation metrics: AUC and F1-score. Figure~\ref{C17-Com} presents a visualization of these metrics for a single MIL model across different feature extractors, using bar charts for both the Camelyon-17-Origin and Camelyon-17-Refine datasets. This visualization effectively illustrates how these metrics vary as the dataset undergoes refinement. Our results indicate that both AUC and F1-score exhibited notable changes following the dataset's adjustment. Figure~\ref{Model-rank} further visualizes the F1-score, AUC, and their combined values, highlighting the top three models to assess the impact of dataset refinement on the performance ranking of the models. While the overall model rankings shifted to some extent after the dataset refinement, the CLAM-MB~\cite{lu2021data} model consistently maintained its top-ranked position, indicating its robustness. 
In summary, dataset refinement enhanced the accuracy of model evaluation metrics and improved the fairness of model rankings, establishing a more solid foundation for future research.

\subsection*{Camelyon\(^+\) Benchmark Experiment}
In the Benchmark Experiment on Camelyon\(^+\), we maintained the same hyperparameter settings as in the comparative experiments on Camelyon-17~\cite{bandi2018detection}.
On the merged Camelyon\(^+\) dataset, we evaluated the MIL approach using feature extractors from two pre-trained natural image models, ResNet-50\cite{he2016deep} and VIT-S\cite{dosovitskiy2020image}, and four pre-trained pathology image models,  PLIP\cite{huang2023visual}, CONCH\cite{lu2024visual}, UNI\cite{chen2024towards}, and Gigapath\cite{xu2024whole}. 
We classify these feature extractors into three main categories: ResNet-50\cite{he2016deep} and ViT-S\cite{dosovitskiy2020image} fall under the domain of natural image pre-training; 
PLIP and CONCH fall under the domain of image-text contrastive learning pre-training; and UNI, and GigaPath belong to the category of pathology-specific visual pre-training. 
We report the mean and variance of model performance in Table~\ref{N IMG}, Table~\ref{WSI Image} and Table~\ref{Text-Img}, which can serve as baselines and references for future work based on the Camelyon\(^+\) dataset.

\begin{figure}[htbp]
\centering
\includegraphics[width=0.96\linewidth]{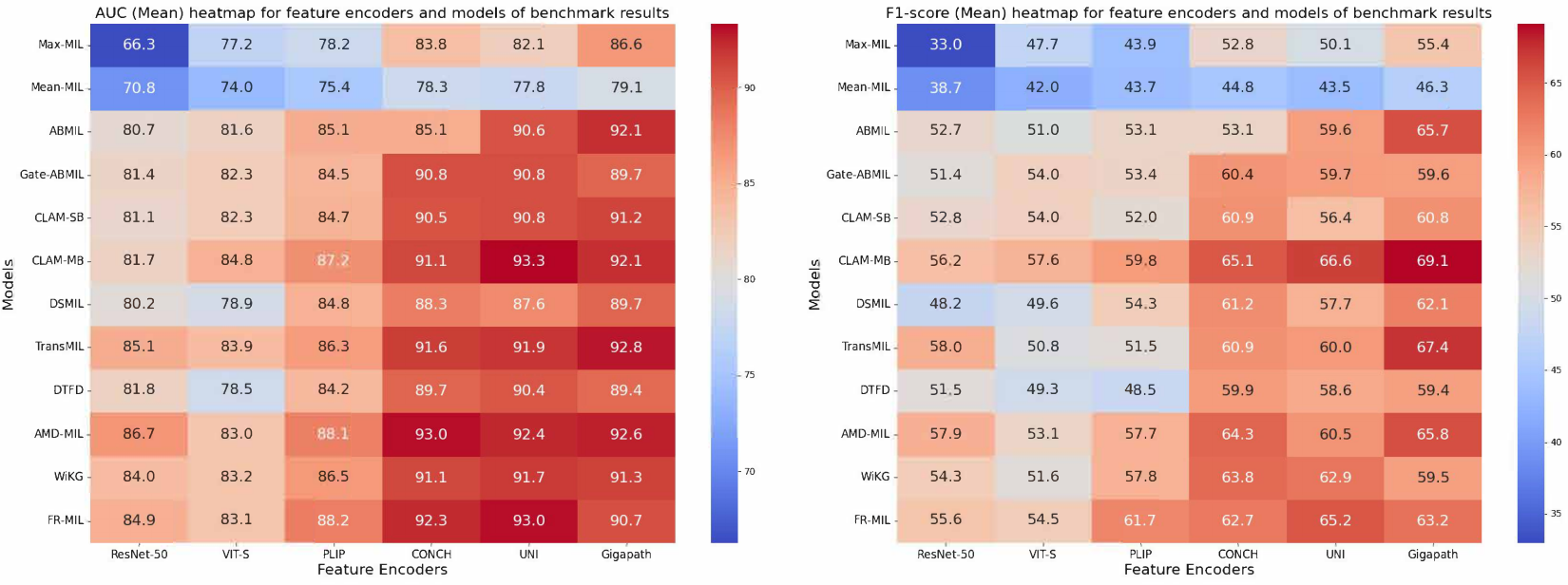}
\caption{Distribution of AUC and F1-score in benchmark results across different feature extractors and aggregators.}
\label{FM}
\end{figure}

As shown in Figure~\ref{FM}, we present a heatmap of the distribution of AUC and F1-score across different MIL models under various feature extractors. It can be observed that pathology-pretrained feature extractors significantly enhance the performance of MIL. Notably, the CONCH\cite{lu2024visual} model, which uses a VIT-Base\cite{ilse2018attention} architecture with image-text contrastive learning, achieves performance comparable to the UNI\cite{chen2024towards} and Gigapath\cite{xu2024whole} models, which utilize VIT-Large\cite{ilse2018attention} and VIT-Giant\cite{ilse2018attention} architectures, respectively. Moreover, both UNI and Gigapath leverage larger training datasets. This suggests that image-text contrastive pretraining may hold greater potential than pure visual pretraining in the pathology domain. While the PLIP~\cite{huang2023visual} model is also pretrained using image-text contrastive learning, its performance does not match that of CONCH, likely due to its smaller dataset and the lower quality of data sourced from Twitter.

\begin{figure}[htbp]
\centering
\includegraphics[width=0.94\linewidth]{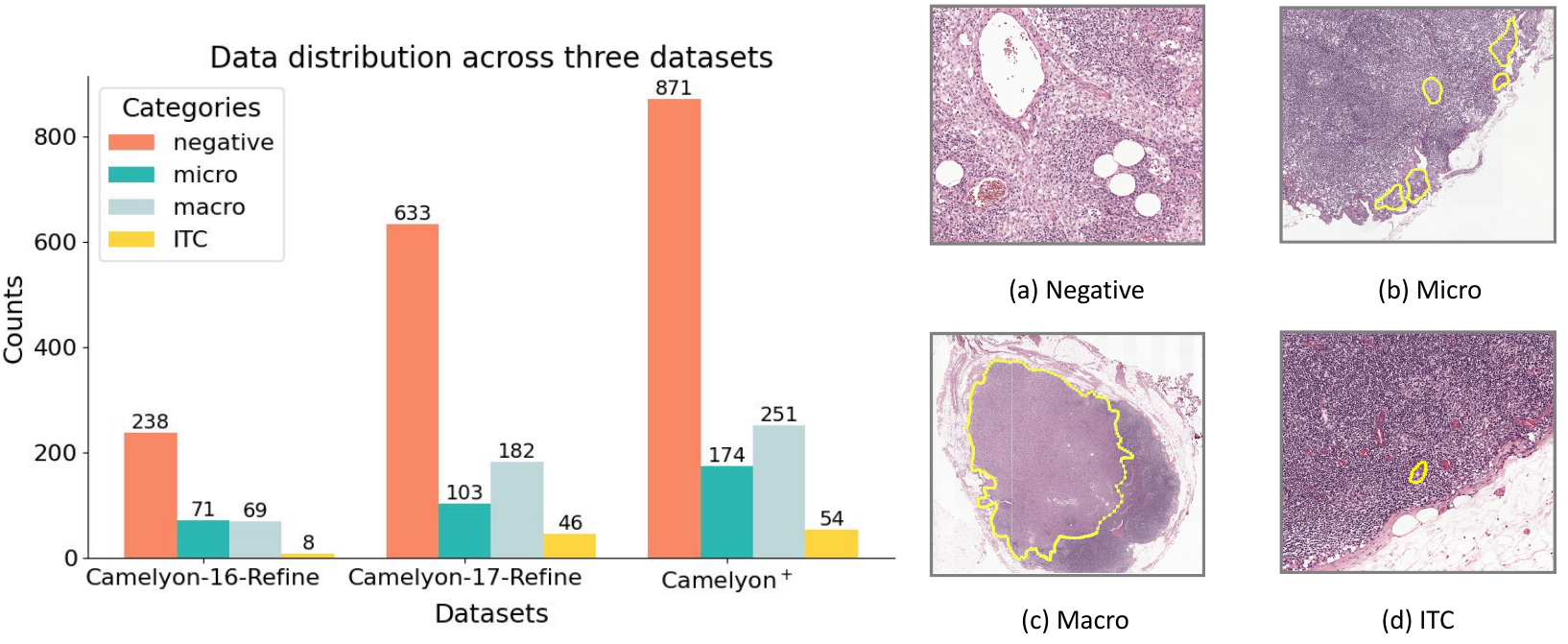}
\caption{Revised data distribution of the Camelyon dataset and the pathological characteristics of different categories.}
\label{DS-QK}
\end{figure}

\begin{figure}[htbp]
\centering
\includegraphics[width=0.96\linewidth]{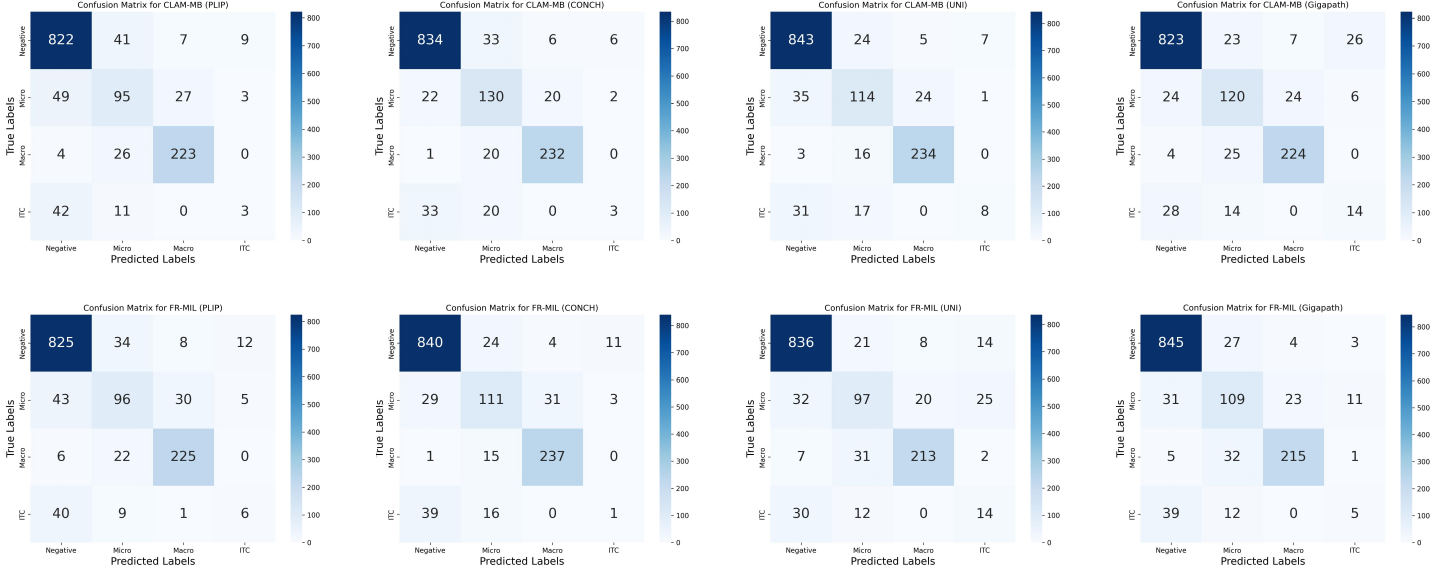}
\caption{Confusion matrices of CLAM-MB~\cite{lu2021data} and FR-MIL~\cite{10640165} under the PLIP~\cite{huang2023visual}, CONCH~\cite{chen2024towards}, UNI~\cite{lu2024visual}, and Gigapath~\cite{xu2024whole} feature encoders.}
\label{HX-JZ}
\end{figure}

In the benchmark results, while the model demonstrates relatively strong performance in terms of accuracy and AUC, the F1-score, recall, and precision are notably low. As illustrated in Figure~\ref{HX-JZ}, we visualized the confusion matrices for the CLAM-MB~\cite{lu2021data} and FR-MIL~\cite{10640165} models. The results show that the models perform relatively well in classifying the negative, micro, and macro categories, but perform poorly in the ITC category. We used macro-averaging to calculate the F1-score, recall, and precision, and the model's poor performance on the ITC category significantly lowered the overall performance metrics.
To investigate this issue further, we analyzed the model's difficulty in identifying ITC cases. One major factor is the severe class imbalance in the Camelyon\(^+\) dataset. As shown in Figure~\ref{DS-QK}, the head class, negative, contains 871 slides, while the tail class, ITC, contains only 54 slides, resulting in an imbalance ratio of approximately 16.1. This imbalance classifies the dataset as having a moderately long-tailed distribution. Such imbalance highlights a key challenge in pathology image analysis: how to achieve balanced model performance on long-tailed datasets like Camelyon\(^+\), particularly since real-world pathological data naturally follow a long-tailed distribution.
Furthermore, we identified a fundamental difference between the four-class classification task in Camelyon\(^+\) and typical cancer subtyping tasks, such as those in TCGA-NSCLC, TCGA-RCC, or BRACS. The ITC, micro, and macro categories in Camelyon\(^+\) are primarily distinguished by the size of metastatic regions, whereas MIL is generally better suited for binary classification tasks, such as detecting the presence or absence of cancer. This explains why models achieve high performance on binary tasks like those in Camelyon-16~\cite{bejnordi2017diagnostic} or Camelyon-17~\cite{bandi2018detection}. Consequently, Camelyon\(^+\) raises an important question about whether the MIL approach the most suitable paradigm for clinical classification tasks like Camelyon\(^+\), where categories are defined by the size of metastatic regions rather than distinct subtypes of cancer.

\subsection*{Evaluation metrics}
In the comparative experiments on the Camelyon-17~\cite{bandi2018detection} dataset and Benchmark experiments, we used accuracy, AUC, F1-score, recall, precision, and kappa value to assess classification performance. The Kappa coefficient is a statistical measure used to evaluate the level of agreement in classification models. It is commonly employed to assess classifier performance, particularly in multi-class classification tasks. This coefficient measures how well the predicted results align with actual labels while accounting for the agreement that could occur by chance. We calculated these metrics using the macro method, while results based on the micro and weighted methods will be available at (\hyperlink{https://github.com/lingxitong/CAMELYON-PLUS-BENCHMARK}{https://github.com/lingxitong/CAMELYON-PLUS-BENCHMARK}).

\section*{Usage Notes}

The Camelon\(^+\) Dataset is publicly available under the Creative Commons Zero (CC0) license. However, please note that this dataset is not intended for developing diagnosis-focused algorithms or models, and should not be used as the sole basis for clinical evaluations in classification tasks.

\section*{Code availability}

The code related to dataset partitioning strategies, hyperparameter configurations, integration of MIL methods, and evaluation metric calculations will be made publicly available at:(\hyperlink{https://github.com/lingxitong/MIL_BASELINE}{https://github.com/lingxitong/MIL\_BASELINE}).

\bibliography{sample}

\section*{Acknowledgements} 

This work was supported by the National Natural Science Foundation of China (NSFC) under Grant No. 82430062. We also gratefully acknowledge the support from the Shenzhen Engineering Research Centre (Grant No. XMHT20230115004) and the Shenzhen Science and Technology Innovation Commission (Grant No. KCXFZ20201221173207022). This work was also supported by the Shenzhen High-level Hospital Construction Fund. Additionally, we thank the Jilin FuyuanGuan Food Group Co., Ltd. for their collaboration.

\section*{Author contributions statement}

X.L. and J.L. conceptualized the study, designed the experiments, and conducted the specific experiments. Y.L., J.C., and W.H. were responsible for dataset correction and construction. T.G., J.G., and Y.H. contributed to the manuscript writing and provided insights into the development of the manuscript structure. All authors read and approved the final version of the manuscript.

\section*{Competing interests} 
The authors declare no competing interests.
\newpage
\section*{Figures \& Tables}

\begin{table}[htbp]
\centering
\caption{Performance metrics of different methods on the Camelyon-17-Origin dataset.}
\begin{tabular}{|c|c|c|c|c|c|c|}
\hline
\textbf{Methods} & \textbf{Acc (\%)} & \textbf{AUC (\%)} & \textbf{F1 (\%)} & \textbf{Recall (\%)} & \textbf{Precision (\%)} & \textbf{Kappa} \\
\hline
\multicolumn{7}{|c|}{PLIP\cite{huang2023visual} (WSIs pre-trained)} \\
\hline
Max-MIL & $83.1 \pm 0.90$ & $79.8 \pm 1.58$ & $46.2 \pm 0.34$ & $45.4 \pm 0.83$ & $46.6 \pm 2.35$ & $0.69 \pm 0.00$ \\
Mean-MIL & $78.7 \pm 0.64$ & $77.1 \pm 0.17$ & $43.1 \pm 0.25$ & $44.5 \pm 0.43$ & $49.3 \pm 1.58$ & $0.56 \pm 0.01$ \\
ABMIL \cite{ilse2018attention} & $84.4 \pm 0.72$ & $86.2 \pm 0.24$ & $56.6 \pm 1.17$ & $54.4 \pm 0.31$ & $53.5 \pm 0.45$ & $0.77 \pm 0.01$ \\
Gate-ABMIL \cite{ilse2018attention} & $83.9 \pm 1.30$ & $86.4 \pm 0.50$ & $55.5 \pm 2.66$ & $53.8 \pm 1.78$ & $53.3 \pm 0.69$ & $0.76 \pm 0.02$ \\
CLAM-SB \cite{lu2021data} & $84.9 \pm 0.50$ & $86.4 \pm 0.32$ & $54.4 \pm 1.63$ & $53.2 \pm 1.05$ & $52.5 \pm 0.55$ & $0.76 \pm 0.01$ \\
CLAM-MB \cite{lu2021data} & $86.1 \pm 0.42$ & $\textbf{89.2} \pm \textbf{0.62}$ & $60.6 \pm 3.30$ & $58.4 \pm 2.58$ & $59.4 \pm 5.25$ & $\textbf{0.79} \pm \textbf{0.02}$ \\
DSMIL \cite{li2021dual} & $86.1 \pm 0.61$ & $87.4 \pm 1.14$ & $57.8 \pm 4.20$ & $57.9 \pm 3.57$ & $\textbf{61.2} \pm \textbf{6.58}$ & $0.76 \pm 0.02$ \\
TransMIL \cite{shao2021transmil} & $85.4 \pm 1.11$ & $88.3 \pm 1.42$ & $61.0 \pm 5.51$ & $58.3 \pm 4.43$ & $56.7 \pm 4.24$ & $0.74 \pm 0.06$ \\
DTFD \cite{zhang2022dtfd} & $85.5 \pm 0.50$ & $86.1 \pm 0.55$ & $52.9 \pm 1.79$ & $52.1 \pm 2.14$ & $52.6 \pm 1.01$ & $0.76 \pm 0.01$ \\
AMD-MIL \cite{ling2024agent} & $83.7 \pm 3.18$ & $88.1 \pm 1.31$ & $58.9 \pm 3.40$ & $56.4 \pm 4.36$ & $55.1 \pm 4.74$ & $0.73 \pm 0.02$ \\
WiKG \cite{li2024dynamic} & $\textbf{86.3} \pm \textbf{1.22}$ & $88.2 \pm 0.41$ & $\textbf{61.2} \pm \textbf{2.91}$ & $\textbf{59.1} \pm \textbf{3.13}$ & $58.4 \pm 3.04$ & $0.78 \pm 0.01$ \\
FR-MIL \cite{10640165} & $82.7 \pm 3.64$ & $84.0 \pm 4.32$ & $52.8 \pm 6.72$ & $52.0 \pm 5.30$ & $52.1 \pm 4.52$ & $0.71 \pm 0.07$ \\

\hline
\multicolumn{7}{|c|}{UNI\cite{chen2024towards} (WSIs pre-trained)} \\
\hline
Max-MIL & $85.3 \pm 0.58$ & $84.1 \pm 1.40$ & $47.2 \pm 0.59$ & $44.8 \pm 0.75$ & $45.0 \pm 4.80$ & $0.72 \pm 0.03$ \\
Mean-MIL & $76.1 \pm 2.66$ & $81.1 \pm 0.55$ & $46.8 \pm 0.74$ & $47.0 \pm 1.25$ & $48.4 \pm 2.37$ & $0.57 \pm 0.03$ \\
ABMIL \cite{ilse2018attention} & $82.4 \pm 0.35$ & $91.3 \pm 0.35$ & $65.4 \pm 1.86$ & $61.2 \pm 1.35$ & $59.7 \pm 1.63$ & $0.73 \pm 0.02$ \\
Gate-ABMIL \cite{ilse2018attention} & $80.9 \pm 1.40$ & $89.9 \pm 0.44$ & $63.6 \pm 5.47$ & $58.8 \pm 4.05$ & $56.7 \pm 3.37$ & $0.71 \pm 0.02$ \\
CLAM-SB \cite{lu2021data} & $81.3 \pm 1.86$ & $91.9 \pm 0.53$ & $66.2 \pm 4.04$ & $59.9 \pm 2.85$ & $58.5 \pm 1.61$ & $0.70 \pm 0.04$ \\
CLAM-MB \cite{lu2021data} & $85.1 \pm 0.31$ & $\textbf{95.0} \pm \textbf{0.41}$ & $\textbf{77.4} \pm \textbf{2.02}$ & $\textbf{68.1} \pm \textbf{1.20}$ & $\textbf{64.9} \pm \textbf{0.55}$ & $0.70 \pm 0.02$ \\
DSMIL \cite{li2021dual} & $83.2 \pm 0.53$ & $92.7 \pm 1.27$ & $67.7 \pm 4.95$ & $61.8 \pm 3.23$ & $60.2 \pm 2.49$ & $0.66 \pm 0.05$ \\
TransMIL \cite{shao2021transmil} & $\textbf{86.0} \pm \textbf{0.35}$ & $93.7 \pm 1.24$ & $69.0 \pm 2.47$ & $64.4 \pm 2.31$ & $64.4 \pm 1.92$ & $\textbf{0.77} \pm \textbf{0.02}$ \\
DTFD \cite{zhang2022dtfd} & $84.4 \pm 0.60$ & $92.7 \pm 0.37$ & $66.6 \pm 5.33$ & $61.7 \pm 5.15$ & $59.2 \pm 4.56$ & $0.77 \pm 0.04$ \\
AMD-MIL \cite{ling2024agent} & $82.5 \pm 1.62$ & $94.0 \pm 0.70$ & $70.2 \pm 1.79$ & $63.0 \pm 1.60$ & $61.1 \pm 1.04$ & $0.63 \pm 0.06$ \\
WiKG \cite{li2024dynamic} & $84.6 \pm 3.20$ & $93.6 \pm 0.87$ & $63.9 \pm 1.87$ & $60.0 \pm 1.72$ & $58.4 \pm 1.25$ & $0.72 \pm 0.08$ \\
FR-MIL \cite{10640165} & $84.1 \pm 0.70$ & $94.3 \pm 0.66$ & $75.0 \pm 3.91$ & $66.2 \pm 2.13$ & $63.5 \pm 1.28$ & $0.71 \pm 0.01$ \\

\hline
\multicolumn{7}{|c|}{Gigapath\cite{xu2024whole} (WSIs pre-trained)} \\
\hline
Max-MIL & $83.5 \pm 1.81$ & $84.7 \pm 3.79$ & $51.5 \pm 5.76$ & $49.5 \pm 5.31$ & $48.2 \pm 4.85$ & $0.74 \pm 0.04$ \\
Mean-MIL & $78.3 \pm 1.36$ & $81.6 \pm 0.17$ & $47.3 \pm 2.31$ & $48.7 \pm 1.69$ & $53.2 \pm 3.11$ & $0.53 \pm 0.05$ \\
ABMIL \cite{ilse2018attention} & $79.2 \pm 2.43$ & $92.1 \pm 0.27$ & $65.7 \pm 6.11$ & $58.1 \pm 5.31$ & $58.6 \pm 0.99$ & $0.68 \pm 0.03$ \\
Gate-ABMIL \cite{ilse2018attention} & $80.7 \pm 1.14$ & $91.6 \pm 0.99$ & $68.0 \pm 1.85$ & $61.9 \pm 1.88$ & $60.0 \pm 1.55$ & $0.72 \pm 0.02$ \\
CLAM-SB \cite{lu2021data} & $80.3 \pm 4.02$ & $92.7 \pm 0.59$ & $66.0 \pm 9.69$ & $60.1 \pm 8.21$ & $61.4 \pm 4.20$ & $\textbf{0.75} \pm \textbf{0.01}$ \\
CLAM-MB \cite{lu2021data} & $85.5 \pm 0.61$ & $\textbf{96.2} \pm \textbf{0.30}$ & $77.3 \pm 1.45$ & $68.3 \pm 0.94$ & $\textbf{65.2} \pm \textbf{0.73}$ & $0.70 \pm 0.02$ \\
DSMIL \cite{li2021dual} & $\textbf{86.2} \pm \textbf{1.71}$ & $93.2 \pm 1.30$ & $70.1 \pm 2.06$ & $65.6 \pm 0.86$ & $63.6 \pm 1.00$ & $0.71 \pm 0.01$ \\
TransMIL \cite{shao2021transmil} & $84.7 \pm 2.21$ & $94.6 \pm 0.36$ & $\textbf{77.4} \pm \textbf{3.45}$ & $\textbf{68.3} \pm \textbf{3.26}$ & $64.6 \pm 3.06$ & $0.73 \pm 0.04$ \\
DTFD \cite{zhang2022dtfd} & $81.5 \pm 1.22$ & $92.0 \pm 0.80$ & $60.8 \pm 3.69$ & $55.6 \pm 3.66$ & $54.2 \pm 2.66$ & $0.75 \pm 0.07$ \\
AMD-MIL \cite{ling2024agent} & $83.5 \pm 1.03$ & $93.9 \pm 0.36$ & $71.3 \pm 0.79$ & $64.7 \pm 0.38$ & $62.9 \pm 1.31$ & $0.74 \pm 0.01$ \\
WiKG \cite{li2024dynamic} & $83.4 \pm 0.72$ & $94.0 \pm 1.13$ & $70.6 \pm 5.09$ & $64.5 \pm 2.23$ & $63.0 \pm 1.46$ & $0.74 \pm 0.05$ \\
FR-MIL \cite{10640165} & $83.9 \pm 0.81$ & $93.7 \pm 1.02$ & $71.8 \pm 5.51$ & $64.8 \pm 3.34$ & $62.3 \pm 2.78$ & $0.70 \pm 0.04$ \\
\hline
\end{tabular}
\label{C17-Origin-Com}
\end{table}

\begin{table}[htbp]
\centering
\caption{Performance metrics of different methods on the Camelyon-17-Refine dataset.}
\begin{tabular}{|c|c|c|c|c|c|c|}
\hline
\textbf{Methods} & \textbf{Acc (\%)} & \textbf{AUC (\%)} & \textbf{F1 (\%)} & \textbf{Recall (\%)} & \textbf{Precision (\%)} & \textbf{Kappa} \\
\hline
\multicolumn{7}{|c|}{PLIP\cite{huang2023visual} (WSIs pre-trained)} \\
\hline
Max-MIL & $77.8 \pm 9.55$ & $72.7 \pm 11.09$ & $38.7 \pm 10.28$ & $37.4 \pm 10.63$ & $41.0 \pm 3.25$ & $0.49 \pm 0.32$ \\
Mean-MIL & $79.8 \pm 0.24$ & $77.7 \pm 0.25$ & $43.2 \pm 0.18$ & $44.8 \pm 0.22$ & $49.2 \pm 0.57$ & $0.59 \pm 0.00$ \\
ABMIL \cite{ilse2018attention} & $81.8 \pm 1.39$ & $87.0 \pm 1.18$ & $52.9 \pm 1.63$ & $51.0 \pm 1.48$ & $51.5 \pm 0.42$ & $0.75 \pm 0.02$ \\
Gate-ABMIL \cite{ilse2018attention} & $81.9 \pm 2.04$ & $87.5 \pm 0.13$ & $53.1 \pm 2.83$ & $51.3 \pm 2.61$ & $51.7 \pm 1.01$ & $0.75 \pm 0.02$ \\
CLAM-SB \cite{lu2021data} & $82.8 \pm 1.73$ & $87.0 \pm 0.76$ & $54.2 \pm 2.32$ & $52.3 \pm 1.95$ & $52.0 \pm 0.88$ & $0.76 \pm 0.02$ \\
CLAM-MB \cite{lu2021data} & $\textbf{87.2} \pm \textbf{0.80}$ & $\textbf{89.3} \pm \textbf{0.41}$ & $61.4 \pm 2.12$ & $\textbf{59.6} \pm \textbf{2.69}$ & $\textbf{61.2} \pm \textbf{7.85}$ & $\textbf{0.82} \pm \textbf{0.01}$ \\
DSMIL \cite{li2021dual} & $86.2 \pm 1.32$ & $87.8 \pm 0.86$ & $56.4 \pm 2.37$ & $56.2 \pm 1.95$ & $56.5 \pm 1.79$ & $0.76 \pm 0.02$ \\
TransMIL \cite{shao2021transmil} & $83.8 \pm 1.00$ & $88.9 \pm 1.15$ & $58.9 \pm 5.70$ & $56.9 \pm 4.14$ & $56.7 \pm 2.78$ & $0.71 \pm 0.02$ \\
DTFD \cite{zhang2022dtfd} & $84.3 \pm 1.81$ & $86.7 \pm 0.95$ & $52.1 \pm 1.79$ & $51.2 \pm 1.12$ & $51.3 \pm 0.19$ & $0.77 \pm 0.01$ \\
AMD-MIL \cite{ling2024agent} & $86.4 \pm 0.97$ & $89.0 \pm 0.67$ & $\textbf{61.9} \pm \textbf{3.01}$ & $59.6 \pm 3.07$ & $58.4 \pm 2.66$ & $0.78 \pm 0.03$ \\
WiKG \cite{li2024dynamic} & $87.1 \pm 1.20$ & $88.2 \pm 1.64$ & $57.9 \pm 2.76$ & $56.6 \pm 2.13$ & $55.8 \pm 1.86$ & $0.80 \pm 0.03$ \\
FR-MIL \cite{10640165} & $80.6 \pm 1.59$ & $87.0 \pm 5.90$ & $57.5 \pm 6.19$ & $55.3 \pm 4.36$ & $54.8 \pm 3.49$ & $0.63 \pm 0.02$ \\

\hline
\multicolumn{7}{|c|}{UNI\cite{chen2024towards} (WSIs pre-trained)} \\
\hline
Max-MIL & $79.6 \pm 10.07$ & $77.9 \pm 11.95$ & $41.3 \pm 10.87$ & $39.1 \pm 9.87$ & $41.1 \pm 2.10$ & $0.55 \pm 0.35$ \\
Mean-MIL & $76.5 \pm 2.02$ & $81.6 \pm 0.26$ & $46.4 \pm 0.70$ & $46.8 \pm 1.09$ & $49.4 \pm 0.83$ & $0.49 \pm 0.06$ \\
ABMIL \cite{ilse2018attention} & $82.1 \pm 1.60$ & $92.3 \pm 0.64$ & $67.7 \pm 1.51$ & $61.9 \pm 1.47$ & $59.6 \pm 1.29$ & $0.72 \pm 0.02$ \\
Gate-ABMIL \cite{ilse2018attention} & $81.0 \pm 0.86$ & $92.2 \pm 0.28$ & $65.7 \pm 4.13$ & $59.7 \pm 3.14$ & $57.9 \pm 2.48$ & $0.68 \pm 0.05$ \\
CLAM-SB \cite{lu2021data} & $81.3 \pm 1.07$ & $93.1 \pm 0.22$ & $64.8 \pm 6.08$ & $58.5 \pm 4.25$ & $56.5 \pm 3.85$ & $0.73 \pm 0.07$ \\
CLAM-MB \cite{lu2021data} & $85.0 \pm 0.68$ & $\textbf{95.9} \pm \textbf{0.21}$ & $74.5 \pm 8.42$ & $65.6 \pm 3.72$ & $63.4 \pm 3.06$ & $0.70 \pm 0.08$ \\
DSMIL \cite{li2021dual} & $85.9 \pm 3.17$ & $93.1 \pm 2.26$ & $62.6 \pm 5.28$ & $59.9 \pm 3.16$ & $59.0 \pm 1.35$ & $0.75 \pm 0.05$ \\
TransMIL \cite{shao2021transmil} & $\textbf{88.5} \pm \textbf{0.44}$ & $95.2 \pm 0.93$ & $70.4 \pm 1.88$ & $65.7 \pm 1.97$ & $66.1 \pm 6.04$ & $0.78 \pm 0.06$ \\
DTFD \cite{zhang2022dtfd} & $82.6 \pm 0.56$ & $94.5 \pm 0.31$ & $61.0 \pm 4.13$ & $56.5 \pm 4.67$ & $56.6 \pm 5.77$ & $\textbf{0.78} \pm \textbf{0.01}$ \\
AMD-MIL \cite{ling2024agent} & $86.0 \pm 1.12$ & $94.8 \pm 0.13$ & $73.6 \pm 3.45$ & $\textbf{68.5} \pm \textbf{1.14}$ & $\textbf{66.8} \pm \textbf{2.55}$ & $0.78 \pm 0.03$ \\
WiKG \cite{li2024dynamic} & $83.1 \pm 3.56$ & $95.0 \pm 0.42$ & $73.3 \pm 1.91$ & $64.6 \pm 3.19$ & $62.6 \pm 1.50$ & $0.68 \pm 0.03$ \\
FR-MIL \cite{10640165} & $85.0 \pm 0.97$ & $96.0 \pm 0.46$ & $\textbf{78.3} \pm \textbf{4.48}$ & $68.0 \pm 1.51$ & $65.4 \pm 1.22$ & $0.70 \pm 0.09$ \\

\hline
\multicolumn{7}{|c|}{Gigapath\cite{xu2024whole} (WSIs pre-trained)} \\
\hline
Max-MIL & $83.7 \pm 2.77$ & $83.4 \pm 3.83$ & $47.3 \pm 0.92$ & $45.9 \pm 0.76$ & $44.9 \pm 1.69$ & $0.75 \pm 0.03$ \\
Mean-MIL & $76.1 \pm 4.24$ & $81.3 \pm 0.55$ & $49.7 \pm 3.29$ & $49.5 \pm 0.91$ & $52.1 \pm 1.32$ & $0.51 \pm 0.06$ \\
ABMIL \cite{ilse2018attention} & $81.4 \pm 0.49$ & $91.8 \pm 0.70$ & $66.6 \pm 2.55$ & $61.6 \pm 2.21$ & $59.8 \pm 2.55$ & $0.75 \pm 0.03$ \\
Gate-ABMIL \cite{ilse2018attention} & $81.4 \pm 1.73$ & $92.7 \pm 0.73$ & $70.6 \pm 1.69$ & $63.2 \pm 2.80$ & $61.1 \pm 2.50$ & $0.72 \pm 0.07$ \\
CLAM-SB \cite{lu2021data} & $78.7 \pm 2.33$ & $92.6 \pm 0.38$ & $59.9 \pm 2.54$ & $54.8 \pm 1.68$ & $55.2 \pm 4.44$ & $0.72 \pm 0.08$ \\
CLAM-MB \cite{lu2021data} & $84.4 \pm 1.92$ & $\textbf{96.5} \pm \textbf{0.44}$ & $\textbf{81.1} \pm \textbf{3.75}$ & $\textbf{68.5} \pm \textbf{0.39}$ & $65.6 \pm 0.30$ & $0.64 \pm 0.09$ \\
DSMIL \cite{li2021dual} & $86.2 \pm 0.85$ & $93.5 \pm 0.61$ & $68.8 \pm 4.85$ & $65.2 \pm 3.73$ & $63.8 \pm 3.42$ & $0.75 \pm 0.06$ \\
TransMIL \cite{shao2021transmil} & $\textbf{86.2} \pm \textbf{0.76}$ & $95.6 \pm 0.36$ & $73.4 \pm 4.51$ & $67.9 \pm 4.69$ & $\textbf{66.5} \pm \textbf{4.56}$ & $\textbf{0.80} \pm \textbf{0.02}$ \\
DTFD \cite{zhang2022dtfd} & $80.9 \pm 1.18$ & $93.4 \pm 0.81$ & $58.1 \pm 2.52$ & $52.7 \pm 1.53$ & $53.5 \pm 1.32$ & $0.76 \pm 0.02$ \\
AMD-MIL \cite{ling2024agent} & $84.6 \pm 1.20$ & $95.0 \pm 0.41$ & $72.9 \pm 1.31$ & $66.4 \pm 1.12$ & $64.9 \pm 2.85$ & $0.75 \pm 0.07$ \\
WiKG \cite{li2024dynamic} & $83.8 \pm 1.41$ & $94.4 \pm 0.52$ & $73.8 \pm 2.83$ & $66.0 \pm 2.31$ & $63.8 \pm 1.39$ & $0.74 \pm 0.04$ \\
FR-MIL \cite{10640165} & $85.9 \pm 0.88$ & $95.4 \pm 0.89$ & $79.7 \pm 6.78$ & $69.2 \pm 2.62$ & $65.7 \pm 1.82$ & $0.72 \pm 0.04$ \\

\hline
\end{tabular}
\label{C17-refine-Com}
\end{table}

\begin{table}[htbp]
\centering
\caption{Benchmark results on Camelyon\(^+\) dastaset of the natural images pre-trained domain.}
\begin{tabular}{|c|c|c|c|c|c|c|}
\hline
\textbf{Methods} & \textbf{Acc (\%)} & \textbf{AUC (\%)} & \textbf{F1 (\%)} & \textbf{Recall (\%)} & \textbf{Precision (\%)} & \textbf{Kappa} \\
\hline
\multicolumn{7}{|c|}{ResNet-50\cite{he2016deep} (ImageNet pre-trained\cite{deng2009imagenet})} \\
\hline
Max-MIL & $70.3 \pm 7.53$ & $66.3 \pm 11.07$ & $33.0 \pm 10.65$ & $28.9 \pm 12.06$ & $30.8 \pm 13.43$ & $0.22 \pm 0.30$ \\
Mean-MIL & $73.2 \pm 2.87$ & $70.8 \pm 4.14$ & $38.7 \pm 3.35$ & $37.9 \pm 2.67$ & $41.1 \pm 1.77$ & $0.39 \pm 0.08$ \\
ABMIL \cite{ilse2018attention} & $81.1 \pm 1.30$ & $80.7 \pm 3.32$ & $52.7 \pm 1.98$ & $51.5 \pm 1.86$ & $52.8 \pm 1.98$ & $0.63 \pm 0.03$ \\
Gate-ABMIL \cite{ilse2018attention} & $80.5 \pm 1.26$ & $81.4 \pm 3.05$ & $51.4 \pm 2.05$ & $49.2 \pm 3.41$ & $51.2 \pm 2.99$ & $0.63 \pm 0.03$ \\
CLAM-SB \cite{lu2021data} & $80.2 \pm 1.78$ & $81.1 \pm 2.93$ & $52.8 \pm 2.07$ & $50.9 \pm 2.03$ & $51.3 \pm 2.51$ & $0.63 \pm 0.02$ \\
CLAM-MB \cite{lu2021data} & $\textbf{83.5} \pm \textbf{1.22}$ & $81.7 \pm 1.89$ & $56.2 \pm 3.80$ & $56.1 \pm 2.83$ & $\textbf{57.9} \pm \textbf{1.75}$ & $\textbf{0.65} \pm \textbf{0.03}$ \\
DSMIL \cite{li2021dual} & $79.8 \pm 1.05$ & $80.2 \pm 4.10$ & $48.2 \pm 0.70$ & $46.9 \pm 1.67$ & $49.1 \pm 4.71$ & $0.61 \pm 0.01$ \\
TransMIL \cite{shao2021transmil} & $81.2 \pm 4.49$ & $85.1 \pm 1.83$ & $58.0 \pm 3.27$ & $\textbf{56.5} \pm \textbf{4.00}$ & $56.8 \pm 5.46$ & $0.63 \pm 0.03$ \\
DTFD \cite{zhang2022dtfd} & $80.6 \pm 1.94$ & $81.8 \pm 4.04$ & $51.5 \pm 2.22$ & $49.7 \pm 2.57$ & $51.2 \pm 3.36$ & $0.63 \pm 0.03$ \\
AMD-MIL \cite{ling2024agent} & $81.1 \pm 1.91$ & $\textbf{86.7} \pm \textbf{3.43}$ & $\textbf{57.9} \pm \textbf{3.03}$ & $55.8 \pm 1.81$ & $55.8 \pm 1.81$ & $0.63 \pm 0.03$ \\
WiKG \cite{li2024dynamic} & $81.3 \pm 3.30$ & $84.0 \pm 3.77$ & $54.3 \pm 3.24$ & $54.3 \pm 3.60$ & $56.2 \pm 3.22$ & $0.62 \pm 0.03$ \\
FR-MIL \cite{10640165} & $82.8 \pm 0.87$ & $84.9 \pm 2.97$ & $55.6 \pm 2.56$ & $56.4 \pm 1.29$ & $58.5 \pm 1.63$ & $0.63 \pm 0.04$ \\
\hline
\multicolumn{7}{|c|}{VIT-S\cite{dosovitskiy2020image} (ImageNet pre-trained\cite{deng2009imagenet})} \\
\hline
Max-MIL & $76.7 \pm 5.30$ & $77.2 \pm 1.90$ & $47.7 \pm 1.17$ & $45.4 \pm 1.97$ & $45.5 \pm 6.72$ & $0.58 \pm 0.10$ \\
Mean-MIL & $72.5 \pm 4.64$ & $74.0 \pm 4.41$ & $42.0 \pm 3.99$ & $41.8 \pm 4.31$ & $45.4 \pm 7.88$ & $0.42 \pm 0.07$ \\
ABMIL \cite{ilse2018attention} & $81.4 \pm 2.07$ & $81.6 \pm 3.62$ & $51.0 \pm 3.26$ & $49.7 \pm 4.06$ & $52.9 \pm 4.71$ & $0.65 \pm 0.04$ \\
Gate-ABMIL \cite{ilse2018attention} & $78.9 \pm 2.05$ & $82.3 \pm 3.41$ & $54.0 \pm 3.10$ & $53.0 \pm 2.78$ & $52.7 \pm 2.55$ & $0.61 \pm 0.05$ \\
CLAM-SB \cite{lu2021data} & $\textbf{81.9} \pm \textbf{1.52}$ & $82.3 \pm 3.57$ & $54.0 \pm 3.25$ & $53.2 \pm 3.30$ & $57.0 \pm 5.80$ & $0.64 \pm 0.06$ \\
CLAM-MB \cite{lu2021data} & $81.7 \pm 3.08$ & $\textbf{84.8} \pm \textbf{4.11}$ & $\textbf{57.6} \pm \textbf{2.34}$ & $\textbf{56.9} \pm \textbf{2.07}$ & $\textbf{59.0} \pm \textbf{3.74}$ & $0.61 \pm 0.09$ \\
DSMIL \cite{li2021dual} & $77.1 \pm 3.38$ & $78.9 \pm 3.28$ & $49.6 \pm 2.83$ & $49.0 \pm 2.75$ & $51.2 \pm 1.73$ & $0.57 \pm 0.07$ \\
TransMIL \cite{shao2021transmil} & $79.7 \pm 1.78$ & $83.9 \pm 2.10$ & $50.8 \pm 4.61$ & $47.1 \pm 5.16$ & $46.7 \pm 7.02$ & $0.63 \pm 0.03$ \\
DTFD \cite{zhang2022dtfd} & $80.4 \pm 1.10$ & $78.5 \pm 2.96$ & $49.3 \pm 2.60$ & $46.4 \pm 3.72$ & $47.4 \pm 7.18$ & $0.63 \pm 0.03$ \\
AMD-MIL \cite{ling2024agent} & $81.7 \pm 0.72$ & $83.0 \pm 2.83$ & $53.1 \pm 3.69$ & $51.6 \pm 3.35$ & $55.1 \pm 3.60$ & $\textbf{0.65} \pm \textbf{0.04}$ \\
WiKG \cite{li2024dynamic} & $80.8 \pm 1.34$ & $83.2 \pm 2.79$ & $51.6 \pm 2.55$ & $50.1 \pm 5.04$ & $51.0 \pm 8.07$ & $0.62 \pm 0.03$ \\
FR-MIL \cite{10640165} & $79.2 \pm 3.24$ & $83.1 \pm 2.03$ & $54.5 \pm 3.56$ & $54.5 \pm 1.93$ & $57.4 \pm 4.86$ & $0.57 \pm 0.06$ \\

\hline
\end{tabular}
\label{N IMG}
\end{table}

\begin{table}[htbp]
\centering
\caption{Benchmark results on Camelyon\(^+\) dastaset of the image-text contrastive pre-trained domain.}
\begin{tabular}{|c|c|c|c|c|c|c|}
\hline
\textbf{Methods} & \textbf{Acc (\%)} & \textbf{AUC (\%)} & \textbf{F1 (\%)} & \textbf{Recall (\%)} & \textbf{Precision (\%)} & \textbf{Kappa} \\
\hline
\multicolumn{7}{|c|}{PLIP\cite{huang2023visual} (WSIs pre-trained)} \\
\hline
Max-MIL & $79.4 \pm 2.49$ & $78.2 \pm 3.44$ & $43.9 \pm 1.87$ & $47.3 \pm 2.13$ & $41.2 \pm 2.01$ & $0.61 \pm 0.03$ \\
Mean-MIL & $73.4 \pm 6.31$ & $75.4 \pm 5.08$ & $43.7 \pm 2.48$ & $43.6 \pm 3.44$ & $48.9 \pm 3.82$ & $0.50 \pm 0.05$ \\
ABMIL \cite{ilse2018attention} & $81.8 \pm 1.94$ & $85.1 \pm 4.15$ & $53.1 \pm 5.62$ & $54.3 \pm 4.08$ & $55.0 \pm 5.96$ & $0.66 \pm 0.03$ \\
Gate-ABMIL \cite{ilse2018attention} & $82.4 \pm 0.98$ & $84.5 \pm 3.91$ & $53.4 \pm 1.57$ & $54.7 \pm 2.14$ & $54.9 \pm 1.68$ & $0.67 \pm 0.04$ \\
CLAM-SB \cite{lu2021data} & $81.9 \pm 1.81$ & $84.7 \pm 4.43$ & $52.0 \pm 6.23$ & $53.8 \pm 4.15$ & $51.8 \pm 8.51$ & $0.65 \pm 0.04$ \\
CLAM-MB \cite{lu2021data} & $83.9 \pm 0.78$ & $87.2 \pm 2.20$ & $59.8 \pm 3.83$ & $60.4 \pm 4.02$ & $60.9 \pm 3.33$ & $0.67 \pm 0.04$ \\
DSMIL \cite{li2021dual} & $81.9 \pm 2.78$ & $84.8 \pm 3.58$ & $54.3 \pm 3.03$ & $54.2 \pm 2.74$ & $56.1 \pm 4.73$ & $0.65 \pm 0.02$ \\
TransMIL \cite{shao2021transmil} & $81.9 \pm 1.57$ & $86.3 \pm 3.01$ & $51.5 \pm 7.23$ & $54.1 \pm 5.52$ & $55.4 \pm 10.71$ & $0.65 \pm 0.04$ \\
DTFD \cite{zhang2022dtfd} & $81.3 \pm 2.22$ & $84.2 \pm 4.23$ & $48.5 \pm 4.74$ & $51.4 \pm 3.52$ & $49.6 \pm 6.55$ & $0.66 \pm 0.03$ \\
AMD-MIL \cite{ling2024agent} & $\textbf{84.9} \pm \textbf{1.45}$ & $88.1 \pm 2.90$ & $57.7 \pm 5.05$ & $59.1 \pm 4.34$ & $58.9 \pm 1.83$ & $\textbf{0.69} \pm \textbf{0.03}$ \\
WiKG \cite{li2024dynamic} & $82.3 \pm 2.34$ & $86.5 \pm 4.16$ & $57.8 \pm 3.53$ & $58.1 \pm 3.31$ & $59.4 \pm 3.46$ & $0.61 \pm 0.08$ \\
FR-MIL \cite{10640165} & $84.6 \pm 0.99$ & $\textbf{88.2} \pm \textbf{3.33}$ & $\textbf{61.7} \pm \textbf{5.41}$ & $\textbf{62.1} \pm \textbf{5.62}$ & $\textbf{63.9} \pm \textbf{7.14}$ & $0.67 \pm 0.05$ \\
\hline
\multicolumn{7}{|c|}{CONCH\cite{lu2024visual} (WSIs pre-trained)} \\
\hline
Max-MIL & $83.8 \pm 3.65$ & $83.8 \pm 5.93$ & $52.8 \pm 8.29$ & $54.0 \pm 6.86$ & $52.8 \pm 10.19$ & $0.66 \pm 0.06$ \\
Mean-MIL & $77.9 \pm 2.86$ & $78.3 \pm 4.54$ & $44.8 \pm 2.23$ & $45.7 \pm 2.65$ & $46.3 \pm 3.95$ & $0.56 \pm 0.04$ \\
ABMIL \cite{ilse2018attention} & $81.8 \pm 1.94$ & $85.1 \pm 4.15$ & $53.1 \pm 5.62$ & $54.3 \pm 4.08$ & $55.0 \pm 5.96$ & $0.66 \pm 0.03$ \\
Gate-ABMIL \cite{ilse2018attention} & $85.7 \pm 1.62$ & $90.8 \pm 4.57$ & $60.4 \pm 3.07$ & $61.4 \pm 3.02$ & $63.0 \pm 3.83$ & $0.70 \pm 0.06$ \\
CLAM-SB \cite{lu2021data} & $86.6 \pm 2.03$ & $90.5 \pm 4.97$ & $60.9 \pm 1.23$ & $61.8 \pm 2.12$ & $64.6 \pm 8.59$ & $0.72 \pm 0.04$ \\
CLAM-MB \cite{lu2021data} & $\textbf{88.0} \pm \textbf{1.95}$ & $91.1 \pm 5.02$ & $\textbf{65.1} \pm \textbf{2.62}$ & $66.6 \pm 2.78$ & $66.2 \pm 6.34$ & $0.73 \pm 0.06$ \\
DSMIL \cite{li2021dual} & $86.8 \pm 1.98$ & $88.3 \pm 4.21$ & $61.2 \pm 1.66$ & $61.5 \pm 1.02$ & $61.8 \pm 3.03$ & $0.70 \pm 0.02$ \\
TransMIL \cite{shao2021transmil} & $84.7 \pm 3.80$ & $91.6 \pm 2.73$ & $60.9 \pm 3.13$ & $62.6 \pm 3.43$ & $64.8 \pm 13.42$ & $0.71 \pm 0.04$ \\
DTFD \cite{zhang2022dtfd} & $86.3 \pm 2.68$ & $89.7 \pm 4.73$ & $59.9 \pm 2.14$ & $61.0 \pm 2.00$ & $59.7 \pm 3.39$ & $0.72 \pm 0.05$ \\
AMD-MIL \cite{ling2024agent} & $84.1 \pm 3.30$ & $\textbf{93.0} \pm \textbf{3.21}$ & $64.3 \pm 6.73$ & $\textbf{66.9} \pm \textbf{8.02}$ & $66.0 \pm 5.91$ & $0.62 \pm 0.17$ \\
WiKG \cite{li2024dynamic} & $86.2 \pm 2.71$ & $91.1 \pm 4.71$ & $63.8 \pm 4.29$ & $65.0 \pm 4.28$ & $\textbf{67.5} \pm \textbf{12.20}$ & $\textbf{0.73} \pm \textbf{0.03}$ \\
FR-MIL \cite{10640165} & $87.3 \pm 2.62$ & $92.3 \pm 3.16$ & $62.7 \pm 3.90$ & $63.7 \pm 5.00$ & $63.1 \pm 2.65$ & $0.70 \pm 0.07$ \\
\hline
\end{tabular}
\label{WSI Image}
\end{table}

\begin{table}[htbp]
\centering
\caption{Benchmark results on Camelyon\(^+\) dastaset of the WSIs vision pre-trained domain.}
\begin{tabular}{|c|c|c|c|c|c|c|}
\hline
\textbf{Methods} & \textbf{Acc (\%)} & \textbf{AUC (\%)} & \textbf{F1 (\%)} & \textbf{Recall (\%)} & \textbf{Precision (\%)} & \textbf{Kappa} \\

\hline
\multicolumn{7}{|c|}{UNI\cite{chen2024towards} (WSIs pre-trained)} \\
\hline
Max-MIL & $81.1 \pm 3.77$ & $82.1 \pm 5.87$ & $50.1 \pm 7.48$ & $52.1 \pm 7.05$ & $50.1 \pm 7.47$ & $0.63 \pm 0.06$ \\
Mean-MIL & $74.1 \pm 4.15$ & $77.8 \pm 3.53$ & $43.5 \pm 4.51$ & $44.4 \pm 3.81$ & $43.8 \pm 5.30$ & $0.49 \pm 0.08$ \\
ABMIL \cite{ilse2018attention} & $84.6 \pm 2.28$ & $90.6 \pm 2.94$ & $59.6 \pm 2.12$ & $60.2 \pm 1.62$ & $60.4 \pm 2.97$ & $0.69 \pm 0.06$ \\
Gate-ABMIL \cite{ilse2018attention} & $84.1 \pm 3.13$ & $90.8 \pm 1.01$ & $59.7 \pm 4.00$ & $60.3 \pm 3.41$ & $63.0 \pm 5.87$ & $0.70 \pm 0.04$ \\
CLAM-SB \cite{lu2021data} & $83.5 \pm 3.41$ & $90.8 \pm 3.82$ & $56.4 \pm 3.59$ & $57.5 \pm 2.91$ & $56.8 \pm 4.38$ & $0.68 \pm 0.03$ \\
CLAM-MB \cite{lu2021data} & $\textbf{88.0} \pm \textbf{1.93}$ & $\textbf{93.3} \pm \textbf{3.40}$ & $\textbf{66.6} \pm \textbf{5.70}$ & $\textbf{67.1} \pm \textbf{6.87}$ & $70.2 \pm 5.74$ & $\textbf{0.75} \pm \textbf{0.04}$ \\
DSMIL \cite{li2021dual} & $83.0 \pm 3.76$ & $87.6 \pm 3.01$ & $57.7 \pm 5.76$ & $56.7 \pm 5.76$ & $64.7 \pm 13.70$ & $0.65 \pm 0.05$ \\
TransMIL \cite{shao2021transmil} & $85.9 \pm 2.32$ & $91.9 \pm 2.59$ & $60.0 \pm 10.56$ & $62.0 \pm 9.04$ & $66.8 \pm 13.03$ & $0.72 \pm 0.03$ \\
DTFD \cite{zhang2022dtfd} & $84.4 \pm 2.97$ & $90.4 \pm 2.93$ & $58.6 \pm 4.08$ & $59.6 \pm 2.76$ & $60.0 \pm 7.16$ & $0.72 \pm 0.05$ \\
AMD-MIL \cite{ling2024agent} & $85.3 \pm 1.76$ & $92.4 \pm 1.19$ & $60.5 \pm 6.29$ & $61.4 \pm 6.37$ & $61.7 \pm 6.59$ & $0.72 \pm 0.03$ \\
WiKG \cite{li2024dynamic} & $85.6 \pm 3.36$ & $91.7 \pm 2.31$ & $62.9 \pm 3.13$ & $64.3 \pm 2.39$ & $64.2 \pm 8.35$ & $0.73 \pm 0.03$ \\
FR-MIL \cite{10640165} & $85.2 \pm 2.81$ & $93.0 \pm 2.20$ & $65.2 \pm 4.46$ & $64.9 \pm 5.57$ & $\textbf{70.9} \pm \textbf{9.53}$ & $0.69 \pm 0.05$ \\

\hline
\multicolumn{7}{|c|}{Gigapath\cite{xu2024whole} (WSIs pre-trained)} \\
\hline
Max-MIL & $81.4 \pm 6.62$ & $86.6 \pm 6.96$ & $55.4 \pm 5.93$ & $54.6 \pm 5.35$ & $54.9 \pm 5.53$ & $0.65 \pm 0.11$ \\
Mean-MIL & $77.4 \pm 3.20$ & $79.1 \pm 3.17$ & $46.3 \pm 4.24$ & $46.3 \pm 4.61$ & $53.6 \pm 9.24$ & $0.53 \pm 0.09$ \\
ABMIL \cite{ilse2018attention} & $79.2 \pm 2.43$ & $92.1 \pm 0.27$ & $65.7 \pm 6.11$ & $58.1 \pm 5.31$ & $58.6 \pm 0.99$ & $0.68 \pm 0.03$ \\
Gate-ABMIL \cite{ilse2018attention} & $84.9 \pm 3.00$ & $89.7 \pm 3.41$ & $59.6 \pm 1.61$ & $58.7 \pm 2.70$ & $59.5 \pm 4.52$ & $0.69 \pm 0.04$ \\
CLAM-SB \cite{lu2021data} & $85.4 \pm 2.93$ & $91.2 \pm 2.47$ & $60.8 \pm 3.78$ & $59.8 \pm 3.56$ & $63.7 \pm 8.90$ & $0.72 \pm 0.03$ \\
CLAM-MB \cite{lu2021data} & $86.7 \pm 3.01$ & $92.1 \pm 2.88$ & $\textbf{69.1} \pm \textbf{8.13}$ & $66.7 \pm 5.36$ & $67.0 \pm 3.40$ & $0.70 \pm 0.07$ \\
DSMIL \cite{li2021dual} & $86.5 \pm 3.12$ & $89.7 \pm 4.70$ & $62.1 \pm 4.48$ & $62.3 \pm 2.86$ & $64.2 \pm 2.46$ & $0.68 \pm 0.06$ \\
TransMIL \cite{shao2021transmil} & $\textbf{88.2} \pm \textbf{2.77}$ & $\textbf{92.8} \pm \textbf{2.44}$ & $67.4 \pm 3.54$ & $\textbf{66.9} \pm \textbf{2.68}$ & $75.5 \pm 11.95$ & $\textbf{0.73} \pm \textbf{0.07}$ \\
DTFD \cite{zhang2022dtfd} & $83.7 \pm 2.33$ & $89.4 \pm 4.64$ & $59.4 \pm 3.78$ & $58.0 \pm 4.24$ & $61.3 \pm 3.57$ & $0.68 \pm 0.07$ \\
AMD-MIL \cite{ling2024agent} & $87.3 \pm 1.92$ & $92.6 \pm 3.11$ & $65.8 \pm 4.68$ & $64.6 \pm 4.28$ & $64.6 \pm 4.42$ & $0.72 \pm 0.05$ \\
WiKG \cite{li2024dynamic} & $85.0 \pm 3.89$ & $91.3 \pm 2.93$ & $59.5 \pm 5.13$ & $59.4 \pm 6.88$ & $68.0 \pm 16.12$ & $0.72 \pm 0.06$ \\
FR-MIL \cite{10640165} & $86.2 \pm 2.01$ & $90.7 \pm 3.11$ & $63.2 \pm 3.54$ & $63.8 \pm 2.40$ & $\textbf{75.7} \pm \textbf{10.20}$ & $0.71 \pm 0.06$ \\

\hline
\end{tabular}
\label{Text-Img}
\end{table}

\end{document}